\documentstyle[12pt,aaspp4]{article}
\begin{document}

\title{Making Clean Energy with a Kerr Black Hole: a Tokamak Model for
Gamma-Ray Bursts}

\author{Li-Xin Li}
\affil{Princeton University Observatory, Princeton, NJ 08544--1001, USA}
\affil{E-mail: lxl@astro.princeton.edu}

\begin{abstract}
In this paper we present a model for making clean energy with a Kerr black 
hole. Consider a Kerr black hole with a dense plasma torus spinning around it.  
A toroidal electric current flows on the surface of the torus, which generates
a poloidal magnetic field outside the torus. On the surface of the tours the 
magnetic field is parallel to the surface. The closed magnetic field lines 
winding around the torus compress and confine the plasma in the torus, as in
the case of tokamaks. Though it is unclear if such a model is stable, we look
into the consequences if the model is stable. If the magnetic field is strong 
enough, the baryonic contamination from the plasma in the torus is greatly 
suppressed by the magnetic confinement and a clean magnetosphere of
electron-positron pairs is built up around the black hole. Since there
are no open magnetic field lines threading the torus and no accretion, the 
power of the torus is zero. If some magnetic field lines threading the black 
hole are open and connect with loads, clean energy can be extracted from the
Kerr black hole by the Blandford-Znajek mechanism.

The model may be relevant to gamma-ray bursts. The energy in the Poynting
flux produced by the Blandford-Znajek mechanism is converted into the kinetic 
energy of the electron-positron pairs in the magnetosphere around the black
hole, which generates two oppositely directed jets of electron-positron pairs
with super-high bulk Lorentz factors. The jets collide and interact with the
interstellar medium, which may produce gamma-ray bursts and the afterglows.   
\end{abstract}

\keywords{black hole physics --- magnetic fields --- gamma rays: bursts}

%\section 1
\section{Introduction}
It is still a mystery what is the central engine for gamma-ray bursts (GRBs).
The central engine should be able to provide not only the enormous amount of
energy (up to $10^{54}$ ergs) for explaining the luminosity and the time 
duration of GRBs, but also
the huge bulk Lorentz factor ($>300$) for explaining the rapid variability
of the luminosity and the non-thermal feature of the spectrum (Piran 1999
and references therein). This requires that the energy produced by the central
engine should be clean of baryonic contamination. The Blandford-Znajek
mechanism associated with stellar mass black holes has been suggested to power
GRBs (Paczy\'nski 1993,1998; M\'esz\'aros \& Rees 1997; Lee, Wijers, \& Brown
2000). With magnetic field lines threading the horizon of a Kerr black hole,
the rotational energy of the black hole can be extracted by the Blandford-Znajek
mechanism (Blandford \& Znajek 1977; Thorne, Price, \& Macdonald 1986). Though
it has been argued that the Blandford-Znajek mechanism can provide the required
enormous energy, it remains unclear if the energy extracted from the black hole
is clean enough for producing the huge bulk Lorentz factor. In fact, when some
open magnetic field lines thread the accretion disk which rotates around the
black hole, the angular momentum and the gravitational binding energy of the
disk can be extracted by the magnetic field lines and an outflow of matter can
be produced (Blandford 1976; Blandford \& Payne 1982). It has been argued that
in the case of a thin accretion disk the power of the disk dominates the power
of the black hole (Blandford \& Znajek 1977; Livio, Ogilvie, \& Pringle 1999;
Li 2000a; for different views see Armitage \& Natarajan 1999, and Lee, Brown,
\& Wijers 2000). The supper-Eddington power of the disk drives a strong baryon 
wind from the disk, thus the energy extracted from the disk is expected to be
full of matter and get mixed up with the energy extracted from the black hole.
Therefore, it remains a question if the Blandford-Znajek mechanism can
produce clean energy.

In this paper we present a model for making clean energy with a Kerr black hole.
Consider a Kerr black hole with a dense plasma torus spinning around it. The
central plane of the torus lies in the equatorial plane of the Kerr black hole. A
toroidal electric current flows on the surface of the torus, which generates a
poloidal magnetic field outside the torus. The electric current flows in the same
direction as the spin of the black hole, so on the rotational axis the magnetic
field is parallel to the spin of the black hole. On the surface of the torus the
magnetic field is parallel to the surface of the torus, and the magnetic pressure
balances the gas pressure of the torus. The magnetic field lines near the
rotational axis thread the horizon of the black hole. See Fig. \ref{fig1} for the
topological structure of the poloidal magnetic field. Suppose the model is
stable and there is no accretion. The huge voltage drop along
the magnetic field lines near the horizon of the black hole, which is induced by
the rotation of the black hole, gives rise to a cascade production of
electron-positron pairs (Blandford \& Znajek 1977). A magnetosphere of
electron-positron pairs is thus built up around the black hole. Due to the small
radii of the plasma particles in the strong magnetic field, the magnetic field
lines winding around the torus prevent the plasma particles from leaving the torus
--- like the magnetic confinement mechanism for tokamaks in controlled nuclear
fusion. The baryonic contamination from the torus is thus greatly suppressed by
the magnetic confinement if the magnetic field is strong enough. Since there are
no open magnetic field lines threading the torus and no accretion, the power of
the torus is zero. If some magnetic field lines threading the horizon of the
black hole are open and connect with loads, clean energy can be extracted
from the black hole by the Blandford-Znajek mechanism.

In the following sections we estimate the time scales relevant to the model
outlined above, and the baryonic contamination from the torus. Based on these
estimations we demonstrate that clean energy can be produced by the
Blandford-Znajek mechanism. Possible applications of the model in producing
GRBs are also discussed.

%\section 2
\section{Estimation of Time Scales}
The poloidal magnetic field in the model outlined above (see Fig. \ref{fig1})
cannot be in a exactly steady state, it will diffuse into the torus\footnote{If
the torus is a type I superconductor, the magnetic field will always be expelled
from the torus by the Meisser effect. However in our model we do not require the
torus to be a superconductor.}. However, if the plasma in the torus is perfectly
conducting, the speed of diffusion for the magnetic field is very slow. To see this,
suppose the plasma in the torus has an electric resistivity caused by Coulomb
collisions (Goldston \& Rutherford 1995)
\begin{eqnarray}
    \eta\approx 7\times10^{-24}\, \ln\Lambda \left({T_p\over 10^{10}
    {\rm K}}\right)^{-3/2}\,{\rm esu}\,,
    \label{tres}
\end{eqnarray}
where $\Lambda$ is a cut-off factor resulting from Debye shielding, $T_p$ is the 
temperature of the plasma. The diffusion time for a magnetic field in a conductor 
is (Jackson 1975)
\begin{eqnarray}
    \tau_D = {4\pi L^2\over c^2\eta}\,,
    \label{dtime}
\end{eqnarray}
where $L$ is the characteristic length of the spatial variation of magnetic
fields, $\eta$ is the resistivity of the conductor, and $c$ is the speed of light.
Inserting equation (\ref{tres}) into equation (\ref{dtime}), we get the diffusion
time for the magnetic field in the torus
\begin{eqnarray}
    \tau_T \approx 10^{12}\, {\rm sec}\,\left({T_p\over 10^{10} {\rm K}}
      \right)^{3/2} \left({L\over 1 {\rm km}}\right)^2\,,
    \label{tt}
\end{eqnarray}
where we have taken $\ln\Lambda \approx 20$. The length $L$ in equation (\ref{tt})
represents the depth that the magnetic field diffuses into the torus. The
diffusion speed of the magnetic field is
\begin{eqnarray}
    v_D = {L\over \tau_T}\approx 10^{-7}\, {\rm cm}\,{\rm sec}^{-1}
      \,\left({T_p\over 10^{10} {\rm K}}\right)^{-3/2} \left({L\over
      1 {\rm km}}\right)^{-1}\,,
    \label{vd}
\end{eqnarray}
which decreases with increasing $L$. Because of the small resistivity of the
plasma, the magnetic field diffuses into the torus very slowly.

Suppose the Kerr black hole is nearly extreme, i.e., $ca/GM_H\approx 1$, where
$M_H$ is the mass of the black hole, $a$ is the specific angular momentum of the
black hole, and $G$ is the gravitational constant. The total energy that can be
extracted by the Blandford-Znajek mechanism is (Li 2000c; see also Okamoto 1992, 
and Lee, Wijers, \& Brown 2000)
\begin{eqnarray}
    E \approx 0.09 M_H c^2 \approx 1.6\times 10^{54}\,{\rm ergs}\,
      \left({M_H\over 10 M_{\odot}}\right)\,.
    \label{ener}
\end{eqnarray}
The power of the Blandford-Znajek Mechanism is (Thorne, Price, \& Macdonald 1986)
\begin{eqnarray}
    P \approx 10^{51}\, {\rm ergs}\,{\rm sec}^{-1}\,
      \left({B\over 10^{15}{\rm Gauss}}\right)^2
      \left({M_H\over 10 M_{\odot}}\right)^2\,,
    \label{powr}
\end{eqnarray}
where $B$ is the strength of the magnetic field on the horizon of the black
hole. For $M_H\approx 10 M_{\odot}$ and $B\approx 10^{15}$ Gauss, $P$ 
greatly exceeds the Eddington limit
\begin{eqnarray}
     L_{\rm Edd} \approx 1.3\times 10^{39}\,{\rm ergs}\,{\rm sec}^{-1}\,
       \left({M_H\over 10M_{\odot}}\right)\,.
\end{eqnarray}
The working time of the Blandford-Znajek mechanism is
\begin{eqnarray}
    \tau_{BZ} = {E\over P}\approx 10^3\, {\rm sec}\,
       \left({B\over 10^{15}{\rm Gauss}}\right)^{-2}
       \left({M_H\over 10 M_{\odot}}\right)^{-1}\,.
    \label{tbz}
\end{eqnarray}
During the time of $\tau_{BZ}$, the depth that the magnetic field diffuses into
the torus is
\begin{eqnarray}
    L \approx 3\, {\rm cm}\,
       \left({B\over 10^{15}{\rm Gauss}}\right)^{-1}
       \left({T_p\over 10^{10} {\rm K}}\right)^{-3/4}
       \left({M_H\over 10 M_{\odot}}\right)^{-1/2}\,,
    \label{bzd}
\end{eqnarray}
which is small compared with the radius of the black hole horizon:
$r_H \approx 10^6\,(M_H/10M_{\odot})$ cm. Thus, during the whole process of the
Blandford-Znajek mechanism, the diffusion of the magnetic field into the torus
can be neglected.

The poloidal electric current for the Blandford-Znajek Mechanism is carried by
the electron-positron pairs in the magnetosphere surrounding the black hole.
The number density of the electron-positron pairs needed by the Blandford-Znajek
mechanism is
\begin{eqnarray}
    n_{ee^+} \approx 10^{16}\,{\rm cm}^{-3}\,\left({B\over 10^{15}{\rm Gauss}
       }\right)\left({M_H\over 10 M_{\odot}}\right)^{-1}\,.
    \label{npair}
\end{eqnarray}
This amount of electron-positron pairs provide enough conductivity along
magnetic field lines, but negligible conductivity across magnetic field lines.
The poloidal electric current flows along magnetic field lines in the 
magnetosphere, crosses magnetic field lines in the load region and inside the 
black hole. Since in our model there are no open magnetic field lines threading
the torus, and no magnetic field lines connecting the black hole with the
torus, the poloidal current does not flow through the torus. In other words,
the torus does not participate in the process of energy extraction and 
transportation.
In our model, the only role played by the torus is to provide the toroidal
current for sustaining the poloidal magnetic field.

In the magnetosphere the motion of charged particles is dominated by the magnetic
field. In other words, the Coulomb collision frequencies are much smaller than
the gyrofrequencies of charged particles. In such a case, the diffusion coefficient
of baryons across the magnetic field is (Goldston \& Rutherford 1995)
\begin{eqnarray}
    D_\perp \approx \nu_c r_L^2\,,
    \label{dperp}
\end{eqnarray}
where $\nu_c$ is the collision frequency of the baryons with the electrons and
the positrons in the magnetosphere, $r_L$ is the Larmor radius of the baryons.
If the baryons are protons, we have (Spitzer 1978)
\begin{eqnarray}
    \nu_c \approx 1\,{\rm sec}^{-1}\,\left({n_{ee^+}\over 10^{16} {\rm
       cm}^{-3}}\right)\left({T_e\over 10^{10} {\rm K}}\right)^{-3/2}\,,
    \label{nu}
\end{eqnarray}
and
\begin{eqnarray}
    r_L \approx 10^{-10}\,{\rm cm}\,\left({T_p\over 10^{10}{\rm K}}\right)^{1/2}
       \left({B\over 10^{15}{\rm Gauss}}\right)^{-1}\,,
    \label{rl}
\end{eqnarray}
where $T_e$ is the effective temperature of the electron-positron gas\footnote{We 
call $T_e$ the effective temperature since the electron-positron pairs in the 
magnetosphere need not to be in thermal equilibrium. $T_e$ is defined by $T_e
\approx m_e v_e^2/k_B$, where $v_e$ is the velocity dispersion of electrons
(positrons), $m_e$ is the mass of electron (positron), and $k_B$ is the Boltzmann
constant.}, $T_p$ is the temperature of the protons in the torus\footnote{Though 
protons are not in thermal equilibrium in the magnetosphere, their velocities can 
still be estimated with $v_p\approx (k_B T_p /m_p)^{1/2}$ since the protons are
assumed to escape from the torus.}, and we have taken $\ln\Lambda\approx 20$ in 
equation (\ref{nu}). From equations (\ref{npair}) - (\ref{rl}), we obtain the 
confinement time for baryons in the torus (i.e. the diffusion time of baryons
across the magnetic field)
\begin{eqnarray}
    \tau_c = {r_H^2\over D_\perp}
      \approx 10^{32}\, {\rm sec}\,\left({B\over 10^{15}{\rm Gauss}}\right)
      \left({T_e\over 10^{10} {\rm K}}\right)^{3/2}\left({T_p\over 10^{10} 
      {\rm K}}\right)^{-1}\left({M_H\over 10 M_{\odot}}\right)^3\,,
    \label{tbaryon}
\end{eqnarray}
which is a huge time scale! (Remember that the lifetime of our universe is only
$\approx 10^{17}$ seconds.) 

The ratio of $\tau_c / \tau_{BZ}$ is
\begin{eqnarray}
    {\tau_c\over \tau_{BZ}}\approx 10^{29}\,\left({B\over 10^{15}{\rm Gauss}}
      \right)^3\left({T_e\over 10^{10} {\rm K}}\right)^{3/2}\left({T_p\over 
      10^{10} {\rm K}}\right)^{-1}\left({M_H\over 10 M_{\odot}}\right)^4\,,
    \label{tbtbz}
\end{eqnarray}
which is a huge number. Therefore, during the Blandford-Znajek process the 
contamination of charged baryons from the torus is negligible.

%\section 3
\section{Clean Energy from the Black Hole}
In our model, all the power of the system comes from the black hole. The power 
of the torus is zero since there are no open magnetic field lines threading the 
torus and no accretion. The only role played by the torus is to provide the
toroidal current for sustaining the poloidal magnetic field. This is an
example that the Blandford-Znajek mechanism
works effectively. Because no material can come out the black hole, the energy
extracted from the black hole is expected to be clean except the contamination
from environments.

A major source of contamination is the plasma in the torus. However, due to the
strong magnetic field that tightly winds around the torus, the contamination by
the charged baryons from the fully ionized torus is greatly reduced. If the 
charged baryons
escape from the surface of the torus at a rate $f_0$ (number of particles per
unit area and per unit time), they will diffuse into the funnel regions above
and below the black hole at a rate
\begin{eqnarray}
    f \approx f_0 e^{-\tau_c/4\tau_{BZ}}\,.
\end{eqnarray}
From equation (\ref{tbtbz}), $e^{-\tau_c/4\tau_{BZ}}$ is a tiny
number with an order of magnitude $10^{-10^{28}}$ for $B \approx 10^{15}$ Gauss,
$M_H \approx 10 M_{\odot}$, and $T_e \approx T_p \approx 10^{10}$ K. Thus, the
contamination from the baryons in the torus is extremely tiny and
completely negligible. 

For a super-strong magnetic field with $B \approx 10^{15}$ Gauss, in a balanced
state the internal pressure of the plasma in the torus is $P \ge B^2/8\pi
\approx 10^{28}$ dyne cm$^{-2}$, which is high enough for neutronization to take
place. In such a state the interior of the torus is dominated by a neutron gas
as in the case of neutron stars. Some of the neutrons may escape from the torus
and enter the funnel regions above and below the black hole, but the amount is not
expected to be much since the neutron gas is enveloped by ions and magnetic fields.
Further more, since the energy exchange between neutrons and electrons/positrons
is very ineffective (Fuller, Pruet, \& Abazajian 2000), the pollution of neutrons
is not a problem here. The neutronization process will also produce a lot of 
neutrinos and anti-neutrinos, the annihilation of them will produce 
electron-positron pairs.

Given the above estimations and arguments, the baryonic contamination from the
torus is completely negligible during the whole process of the Blandford-Znajek
mechanism. If the system starts with a clean environment (suppose all baryonic
particles either fall into the black hole or sit in the torus at the beginning), 
then the environment (at least in the funnel regions above and below the black 
hole) will be kept clean from the contamination by the baryonic particles from 
the torus. Then the content in the magnetosphere is overwhelmingly dominated by 
the electron-positron pairs in both number and mass. The mass density of the 
electron-positron pairs in the magnetosphere is
\begin{eqnarray}
    \rho_{ee^+} \approx 2 m_e n_{ee^+} \approx
       10^{-11}\,{\rm g}\,{\rm cm}^{-3}\,
       \left({B\over 10^{15}{\rm Gauss}}\right)
       \left({M_H\over 10 M_{\odot}}\right)^{-1}\,,
    \label{mee}
\end{eqnarray}
where equation (\ref{npair}) has been used. The mass density of the baryonic 
particles in the interstellar medium is
\begin{eqnarray}
    \rho_b \approx 10^{-19}\,{\rm g}\,{\rm cm}^{-3}\,
       \left({n_b\over 10^5{\rm cm}^{-3}}\right)\,,
    \label{mb}
\end{eqnarray}
where $n_b$ is the number density of the baryonic particles in the interstellar
medium, which is $\approx 10^3 - 10^5 \,{\rm cm}^{-3}$ in the regions of ionized 
gas around recently formed massive stars (Garay \& Lizano 1999). Since $\rho_b$ 
is only a tiny fraction ($\sim 10^{-8}$) of $\rho_{ee^+}$, the baryonic 
contamination from the interstellar medium is also negligible.

It is far from clear where the loads are and how the energy of the Poynting 
flux produced by the Blandford-Znajek mechanism is converted into the kinetic 
energy of plasma particles (Thorne, Price, \& Macdonald 1986; Li 2000b). 
The consideration of the instability of the magnetic field
indicates that the loads cannot be far from the black hole and the energy 
extracted from the black hole by the Blandford-Znajek mechanism should be 
converted into the kinetic energy of plasma particles close to the black hole 
(Li 2000b). Suppose the energy extracted from the black hole is converted into 
the kinetic energy of the plasma particles in the magnetosphere to produce an 
outflow of electron-positron pairs. The mass flow is
\begin{eqnarray}
    \dot{M} \approx \rho_{ee^+} Ac \approx 10^{16}\,{\rm g}\,{\rm sec}^{-1}\,
    \left({B\over 10^{15}{\rm Gauss}}\right)
    \left({M_H\over 10 M_{\odot}}\right)
    \left({A\over 10^4 r_H^2}\right)\,,
    \label{mflow}
\end{eqnarray}
where $A$ is the cross-sectional area of the flow. If the efficiency in
converting Poynting flux into the kinetic energy is $\epsilon$, the bulk
Lorentz factor of the outflow is
\begin{eqnarray}
    \Gamma = {P\over \dot{M} c^2} \approx 10^{12}\, 
        \left({\epsilon\over 0.01}\right)
        \left({B\over 10^{15}{\rm Gauss}}\right)\left({M_H\over 10 M_{\odot}}
	\right)\left({A\over 10^4 r_H^2}\right)^{-1}\,,
    \label{gamma}
\end{eqnarray}
where equation (\ref{powr}) and equation (\ref{mflow}) have been 
used\footnote{Equation (\ref{npair}) gives a lower limit on the number density
of electron-positron pairs in the magnetosphere. The number density of 
electron-positron pairs in the magnetosphere may be greater than that given in
equation (\ref{npair}), then the Lorentz factor of the outflow is somewhat 
smaller than that given in equation (\ref{gamma}).}. Thus, the black hole produces 
two oppositely directed jets of electron-positron pairs with super-high bulk 
Lorentz factors along the rotation axis of the black hole. The jets cannot 
produce GRBs by themselves since they have too
high Lorentz factors and too few baryons. However, the jets will collide and 
interact with the interstellar medium. The interaction between the jets and the 
interstellar medium will load more baryons to the jets and decelerate the jets, 
which may produce GRBs and the afterglows.

%\section 4
\section{Discussions}
The magnetic configuration in the model is reasonable in physics, but a real
case may not be so simple. The toroidal current may distribute in the entire 
space inside the torus, not just on the surface. In such a case, a poloidal 
magnetic field will also exist inside the torus. But we can still imagine that 
the surface of the torus coincides with a toroidal magnetic surface\footnote{A 
magnetic surface is a surface of constant magnetic flux.}, so that on the surface
of the torus the magnetic field is parallel to the surface and outside the torus 
the poloidal magnetic field is still like that in Fig. \ref{fig1}. Inside the 
torus the magnetic surfaces are nested toroidal surfaces. This is a very natural 
configuration of the magnetic field generated by a toroidal electric current 
flowing in the torus. Because of the differential rotation of the torus, the 
plasma and the magnetic field inside the torus are subject to the Balbus-Hawley 
instability (Balbus \& Hawley 1991, 1998). Within a few rotational periods the
magnetic field lines inside the torus become chaotic and tangled, the torus
becomes turbulent. If there is no magnetic field outside the torus and the
plasma density outside the torus is low, the boundary of the torus is free and
the turbulence inside the torus will make the torus expand and lead to strong
accretion (Hawley 2000). However, in our model there is a hard ``magnetic wall''
surrounding the torus, which confines the plasma in the torus so that the
boundary of the torus is not free. The ``magnetic wall'' may suppress the 
accretion and the expansion. Thus, the effects of the Balbus-Hawley instability 
inside the torus will just be to redistribute the 
angular momentum and heat the plasma in the torus. The transfer of angular 
momentum within the torus caused by the Balbus-Hawley 
instability will spread the torus in radial direction.
The heat generated by the Balbus-Hawley instability may be rapidly radiated 
away by neutrinos. However, these do not affect the topology of the magnetic
field outside the torus. Since the magnetosphere has a very low mass density, the
magnetic field outside the torus does not suffer from the Balbus-Hawley
instability. No matter how complex the magnetic field inside the torus is, the
magnetic field outside the torus is ordered and  has the topology shown in Fig.
\ref{fig1}. The magnetic field inside the torus is unimportant for our current
purposes, it is the magnetic field outside the torus that is relevant to the 
Blandford-Znajek mechanism and the magnetic confinement. If we only consider the 
magnetic field outside the torus, it is always equivalent to assuming an electric 
current flowing on the surface of the torus. Thus, the magnetic field structure 
shown in Fig. \ref{fig1} is quite general even for real cases.

All the requirements for our model are a Kerr black hole with a dense torus, and
a toroidal electric current flowing in the torus. A Kerr black hole with a dense
torus spinning around it may form from the merger of two neutron star, or the
merger of a black hole and a neutron star, or directly from the core collapse of 
a rapidly rotating massive star (Narayan, Paczy\'nski, \& Piran 1992; Woosley 
1993). But it is quite unclear how the current flows in the torus. Let's
consider the collapse of a magnetized and rapidly rotating massive star. If the 
magnetic field of the star is dipole-like, the magnetic field lines near the 
equatorial region are closed, while those in the polar cap region are open.
If the star has too much angular momentum, a dense torus will be left behind
and rotates around the Kerr black hole that is formed by the core collapse
(Woosley 1993). Because of the different strength of the centrifugal force, the
stellar material in the polar cap region collapses into the Kerr black hole but
the stellar mantle near the equatorial region is left in the torus. It is
possible that all open magnetic field lines are trapped on the horizon of the
black hole, only closed magnetic field lines are left in and around the torus.
At the beginning (i.e. the time right after the formation of the torus) the
magnetic field may have a very complicated structure, but after the torus
settles in a somewhat steady state it can be expected that the magnetic field is
equivalent to be generated by a toroidal current flowing in the torus. The
Balbus-Hawley instability may make the magnetic field inside the torus very
complicated, but the magnetic field outside the torus is stable and is
effectively generated by a toroidal current flowing on the surface of the
torus. This provides a possible astronomical scenario which produces a magnetic
field as that shown in Fig. \ref{fig1}. As another (probably more ideal
and extreme) example, assume that the distribution of the magnetic field in the
star is so non-uniform that right after the core collapse all the magnetic field
lines are trapped on the horizon of the Kerr black hole, none are left in the
torus. This produces such an initial state: a non-magnetized torus rotating
around a highly magnetized Kerr black hole. But this state cannot be kept for a
long time since the black hole has no steady internal current to sustain the
magnetic field. A black hole is a poor conductor, a magnetic field diffuses
away from the horizon of a black hole at the speed of light (Thorne, Price, \&
Macdonald 1986). Thus the magnetic field lines leave the black hole quickly.
The change in the magnetic flux through the horizon of the black hole induces 
a transient toroidal current on the horizon, which causes energy dissipation
and slows down the black hole. But the effect is not significant since the energy
stored in the magnetic field is much smaller than the vast rotational energy 
of the black hole. The expanding magnetic field lines will finally hit the
torus that rotates around the black hole.  Since the plasma in the torus has very
small resistivity, it will be very difficult for the magnetic field to diffuse into
the torus. In stead, the magnetic field lines will bend over the torus, close on
themselves, and a strong toroidal current will be induced on the surface of the
torus. The magnetic field lines embracing the torus will compress the torus and
enhance the temperature and the gas pressure inside the torus, until a balanced
state is reached. In the balanced state the pressure of the plasma in the torus 
balances the pressure of the magnetic field at the surface of the torus, and the 
electric current at the surface of the torus sustains the magnetic field of the 
system. This magnetic field formed in this scenario is just like that shown in Fig. 
\ref{fig1}.

The model is more or less similar to the tokamak in controlled nuclear fusion: 
they both use magnetic fields to confine plasma. It is well-known that both tori
and tokamaks suffer from various hydrodynamical and megnetohydrodynamical
instabilities (Balbus \& Hawley 1998; Bateman 1978). It seems no doubt that such
instabilities must also exist in the torus in our model, but it is unclear what
their consequences are. An instability may be so strong that the global 
equilibrium structure of the model is destroyed, then the model will not work.
On the other hand, an instability that saturates at a modest amplitude may have
little effect. It is also possible that an instability acts as a significant
mechanism for transporting angular momentum in the torus but does not destroy
the global equilibrium state (Balbus \& Hawley 1998). The issue of stabilities 
is so complex that it is beyond the scope of the paper. Thus, in this paper 
we don't try to test the instabilities. Instead, assuming the instabilities are 
not strong enough to destroy the global configuration in the model, we look 
into the consequences of the model.

%\section 5
\section{Conclusions}
We have presented a model for making clean energy with a Kerr black hole.
Assuming a toroidal current flows on the surface of the torus spinning around
a Kerr black hole, a poloidal magnetic field is generated outside the torus. 
The poloidal magnetic field winds around the torus, compresses and confines 
the torus in a way similar to tokamaks for controlled nuclear fusion. Assuming
the torus is stable and has no accretion, a clean magnetosphere of 
electron-positron pairs is built up around the black hole. The baryonic 
contamination from the torus is greatly suppressed by
the magnetic confinement if the magnetic field is strong enough. Since there are
no open magnetic field lines threading
the torus\footnote{We emphasize that there are no {\it open} magnetic field lines
threading the torus. There could be some closed magnetic field lines threading the
torus, it doesn't matter since the closed magnetic field lines can only transfer
angular momentum within the torus.} and no accretion, the power of the torus is
zero. This provides an ideal environment for the working of the Blandford-Znajek
mechanism. If some magnetic field lines threading the horizon of the Kerr black
hole are open and connect with loads, clean energy can be extracted from the
black hole by the Blandford-Znajek mechanism.

In a real case the toroidal electric current may flow in the entire space inside
the torus, not just on the surface. However, if we only consider the magnetic
field outside the torus, the magnetic field is effectively generated by a
toroidal current flowing on the surface of the torus, as sketched in Fig. 
\ref{fig1}. In our model we have
neglected various dynamical instabilities. The instabilities are important
and may make the model unstable, but we look into the consequences of the model
if it is stable.

The energy extracted by the Blandford-Znajek mechanism comes out of the black hole
in the form of Poynting flux. If the energy in the Poynting flux is converted into
the kinetic energy of the plasma particles in the magnetosphere which is dominated
by electron-positron pairs, two oppositely directed jets of electron-positron pairs
with super-high Lorentz factors given by equation (\ref{gamma}) will be produced
along the rotation axis of the black hole. The jets will collide and interact with 
the interstellar medium, which may produce GRBs and the afterglows. This suggests
a tokamak model for GRBs.

In the traditional model of a Kerr black hole with a thin disk, the toroidal
current flowing in the disk generates a poloidal magnetic field threading the
horizon of the black hole as well as the disk. If some magnetic field lines
threading the disk are open, energy and angular momentum can be extracted from
the disk, which drives a strong baryon wind from the disk (Blandford 1976;
Blandford \& Payne 1982). At large distance from the black hole and the disk,
the magnetic field becomes weak and the inertia of particles becomes important.
In fact, in the load region particles must flow across magnetic field lines to
close the poloidal electric current.
Thus, it can be expected that at large distance the baryon wind from the disk 
will get mixed with the clean energy extracted from the black hole, though close 
to the black hole they may be separated from each other. In other words, for the 
traditional model, one may expect clean energy to flow along the magnetic field 
lines threading the black hole only if the baryons lost from the disk do not mix
in. While in our model, the baryons are prevented from leaving the torus since 
there are no open magnetic field lines threading the torus.

The main purpose of the paper is to present a novel geometry of magnetic
fields for a popular object: a spinning black hole and a magnetized torus
around it. While it is far from clear if such a configuration can be formed 
in a realistic astrophysical context, and if it can be stable for a long 
enough time to allow a significant extraction of the spin energy of the black 
hole, we present it here as it has a nice property: it offers a possibility for
extracting clean energy with no baryonic contamination. Probably the specific 
configuration of the magnetic field in the model can only
exist in some special circumstances. However, we should remember that GRBs are 
very rare --- about $10^4$ times less common than supernovae (Paczy\'nski 1999).

\acknowledgments{I am very grateful to Bohdan Paczy\'nski for many discussions and
suggestions. This work was supported by the NASA grant NAG5-7016.}

%REFERENCES

\newpage
%Figure Captions
\figcaption[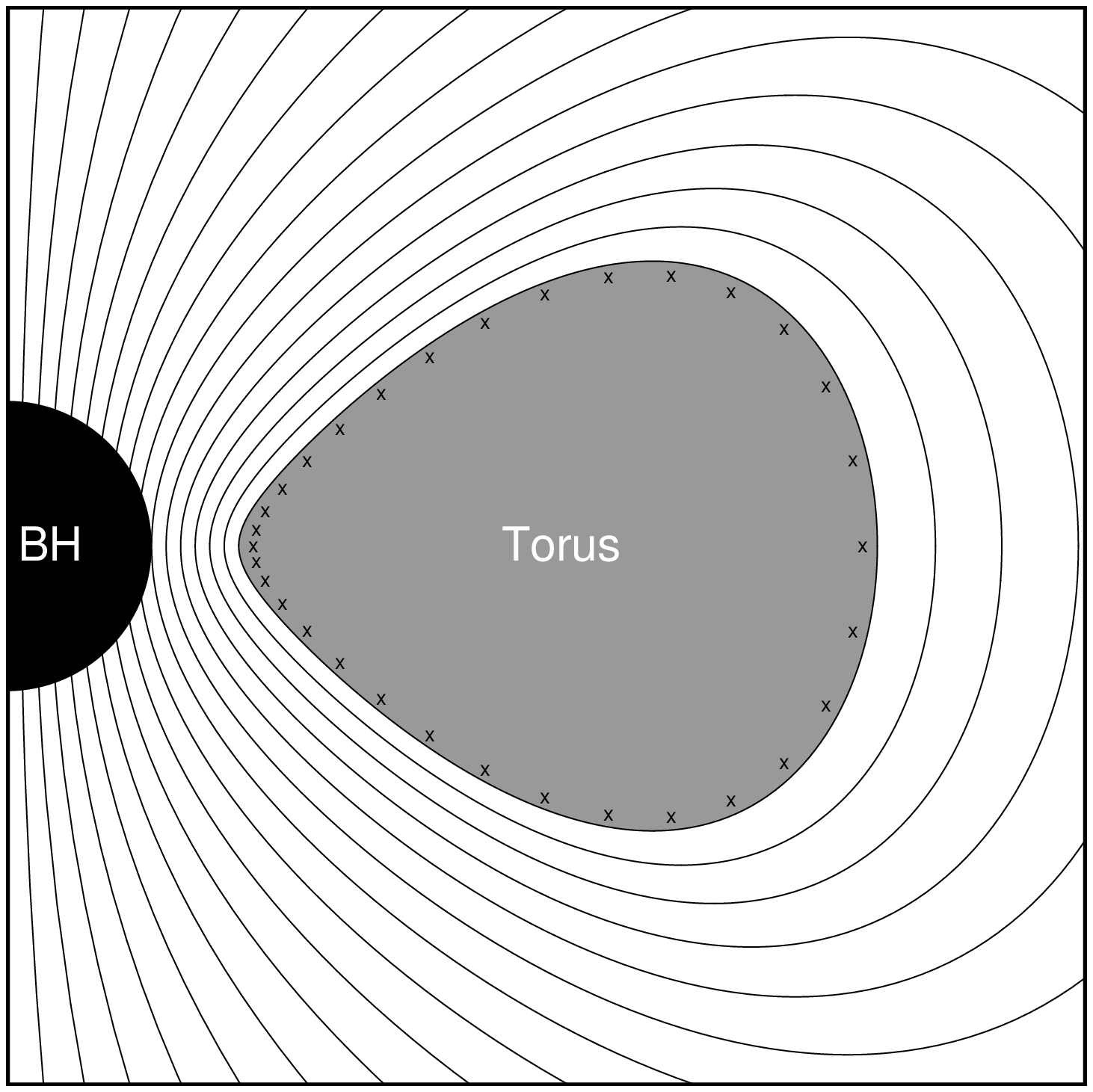]{A cross-sectional diagram for the topological structure 
of the poloidal magnetic field in the model for making clean energy with a Kerr 
black hole. The dark region marked with ``BH'' is a Kerr black hole, the gray
region marked with ``Torus'' is a cross-section of the torus that rotates
around the black hole. The lines outside the torus are the poloidal magnetic 
field lines. On the surface of the torus the magnetic field is parallel to the 
surface. The poloidal magnetic field is generated by a toroidal electric current 
flowing on the surface of the torus, which is shown with the cross signs in the 
gray region. The three-dimensional view is obtained by rotating the configuration 
$360^{\circ}$ around the axis of the black hole (i.e. the left edge of the 
diagram). Baryonic particles are confined in the torus by the strong magnetic
field winding around the torus, so a space clean of baryons exists outside the
torus. Since there are no open magnetic field lines threading the torus and no
accretion, the power of the torus is zero. If some magnetic field lines threading 
the horizon of the black hole are open and connect with loads, clean energy 
can be extracted from the black hole by the Blandford-Znajek mechanism.
\label{fig1}}

\end{document}